# On-demand generation of indistinguishable polarization-entangled photon pairs

M. Müller[†], S. Bounouar[†\*], K. D. Jöns and P. Michler[\*]

August 21, 2013

*Institut für Halbleiteroptik und Funktionelle Grenzflächen, Universität Stuttgart, Allmandring 3, 70569 Stuttgart, Germany.*
[†]*These authors contributed equally to this work.* [\*]*e-mail: s.bounouar@ihfg.uni-stuttgart.de; p.michler@ihfg.uni-stuttgart.de*

**An on-demand source of indistinguishable and entangled photon pairs is a fundamental component for different quantum information applications such as optical quantum computing, quantum repeaters, quantum teleportation and quantum communication [1]. Parametric down-conversion [2,3] and four-wave mixing sources [4] of entangled photons have shown high degrees of entanglement and indistinguishability but the probabilistic nature of their generation process also creates zero or multiple photon pairs following a Poissonian distribution. This limits their use in complex algorithms where many qubits and gate operations are required. Here we show simultaneously ultra-high purity ($g^{(2)}(0) < 0.004$), high entanglement fidelity ($0.81 \pm 0.02$), high two-photon interference non-post selective visibilities ($0.86 \pm 0.03$ and $0.71 \pm 0.04$) and on-demand generation of polarization-entangled photon pairs from a single semiconductor quantum dot (QD). Through a two-photon resonant excitation scheme, the biexciton population is deterministically prepared by a $\pi$-pulse. Applied on a quantum dot showing no exciton fine structure splitting, this results in the deterministic generation of indistinguishable entangled photon pairs.**

To date, spontaneous parametric down-conversion (SPDC) and four wave mixing sources have been mostly used for the generation of entangled photon pairs to realize quantum communication protocols and to demonstrate basic quantum logic experiments [5]. However, the photon pair statistics of these sources is described by a Poissonian distribution which implies also the generation of zero and multiple pairs. This leads to errors in the quantum algorithm protocols [6] which effectively limit their usefulness for deterministic quantum technologies. Radiative cascades in single quantum emitters, such as atoms [7] or quantum dots [8], can in principle emit on demand single pairs of polarization-entangled photons with high generation efficiencies [9]. After optical excitation of two electron-hole pairs (biexciton, called $|XX\rangle$ state) in a quantum dot, the biexciton decays through a two-photon cascade (Fig. 1a). If the fine structure splitting between the intermediate states (excitons called $|X\rangle$) is smaller than the radiative linewidth, the two decay paths are indistinguishable and the two photons are polarization-entangled which results in a two-photon Bell state $|\psi^+\rangle = \frac{1}{\sqrt{2}}(|H_{XX}\rangle|H_X\rangle + |V_{XX}\rangle|V_X\rangle)$. To ensure the emission of a single pair of entangled photons per excitation pulse the biexcitonic state has to be pumped into saturation. So far, non-resonant pulsed pumping schemes have been successfully applied for entangled photon generation [10,11] but no simultaneous information on indistinguishability has been provided. Anyhow, it is well known that non-resonant pumping schemes limit the coherence and indistinguishability of the emitted photons making them unfeasible for many quantum information applications. In a recent study, Stevenson and co-workers reported interference and entanglement properties of photons emitted by a QD embedded within a light-emitting diode [12]. However, the measurements have been performed in a continuous and sinusoidal ac mode without clear pulse separation, and a post-selected visibility of 0.6 was achieved which was limited by detector jitter. Resonant coherent excitation on quantum dots has proven to be very efficient for achieving a high degree of indistinguishability and coherence [13]. On-demand single-photon sources with a high degree of indistinguishability of 97 percent have been recently realized after coherent resonant $\pi$-pulse excitation of an excitonic state in a QD [14]. However, implementation of one-photon resonant excitation of the biexcitonic state is not possible due to the optical selection rules. Here, a resonant two-photon excitation (TPE) scheme is applied to coherently populate the biexcitonic state [15–17]. By setting the pump intensity so that the inversion of the QD from the ground to the biexcitonic state is most probable (so called $\pi-$pulse), one can deterministically prepare a biexciton after each pulse with near unity fidelity. This gain of efficiency is achieved without any drawback on the multi-photon event probability as it is intrinsically the case for SPDC or typically reported for incoherent pumping schemes of QDs due to increased contribution of spectrally nearby transitions [18]. Even more important, this also strongly attenuates the contribution to decoherence which affects solid state emitters since no phonon relaxation processes are needed for the biexciton preparation and no charge carrier is released in the semiconductor matrix preventing carrier-carrier scattering processes with nearby charge carriers.

We used an epitaxially grown (In,Ga)/GaAs QD for our studies (see Method section). Figure 1b shows a photoluminescence spectrum under above bandgap excitation. The exciton (1.4212 eV) and biexciton lines (1.4189 eV) are separated by the biexciton binding energy (2.3 meV). The higher energy line corresponds to the trion, constituted of an exciton and an excedentary charge. Previous high-resolution polarization dependent studies [11] showed that the exciton fine structure splitting is below 1 $\mu$eV. In the following, we address the quantum dot by tuning a linearly polarized pulsed laser (FWHM = 95 $\mu$eV) in resonance to the virtual state of the biexciton two-photon excitation (see Fig. 1a). Fig. 1c shows a QD spectrum under this two-photon excitation. The intensities of the two lines are similar since only a biexciton can be prepared after each pulse, and the exciton is populated through the radiative cascade after the recombination of the biexciton. Since the energy of the laser pump differs by half of the biexciton binding energy from the exciton and biexciton lines, a laser background free signal of both lines (X and XX) can be collected without any polarization suppression technique typically used for resonant excitation experiments on the exciton [19,20]. On one hand, this allows polarization dependent cross-correlation measurements of the biexciton-exciton cascade (see below), and on the other hand perfect single-photon purity ($g^{(2)}(0) < 0.004$) is achieved for both transitions (see Fig. 2).

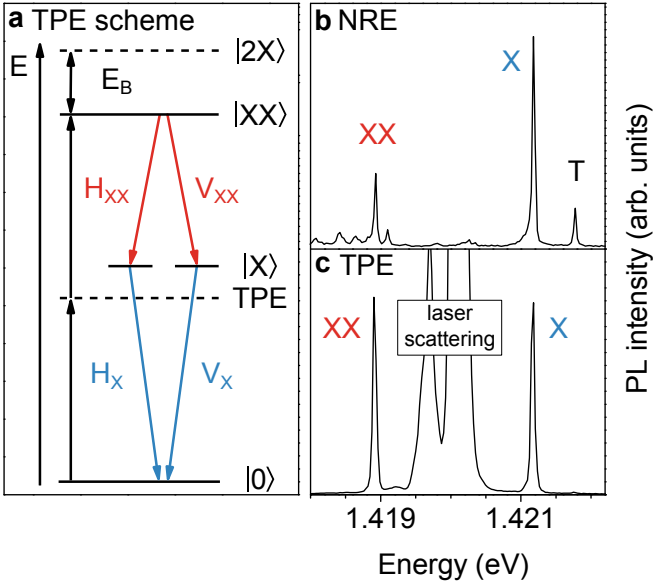

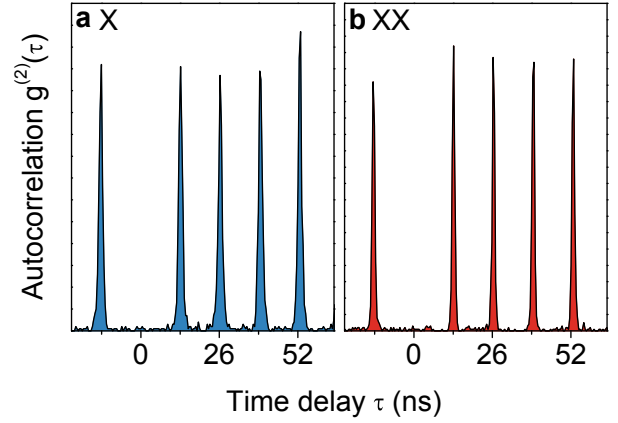

**Figure 1** | **a,** Two-photon biexciton excitation scheme for a quantum dot exhibiting no excitonic fine structure splitting. A shaped laser is brought to resonance with the virtual biexciton two-photon excitation state, which energy is situated between the exciton and the biexciton emission lines. Two radiative recombination paths are possible to the ground state $|0\rangle$ via one of two bright exciton states $|X\rangle$. The polarizations of the two photons are determined by the intermediate exciton state. **b,** Above GaAs bandgap excitation spectrum. The presence of the trion is due to charge capture in the quantum dot around the same time as the exciton or after the biexciton recombination. **c,** Emission spectrum under biexciton direct two-photon excitation: Only the exciton and the biexciton can be observed with same intensities. No trion line is observed. This confirms that exciton and biexciton photons are always emitted by pairs and that the exciton population can not be transferred to the trion state by capture of a charge before the radiative cascade is over. The large peak between the exciton and the biexciton emission lines is due to some residual scattered laser light.

Figure 3 shows the excitation power dependence of the integrated intensity of the exciton and biexciton lines, respectively. Well pronounced Rabi oscillations are observed for the biexciton [15,16] emission intensities (Fig. 3a). They reflect the coherent nature of the two-photon excitation process together with a high biexciton occupation probability. The observed damping is due to decoherence processes which might be caused by an enhanced coupling to phonons when the quantum dot state rotates faster between $|0\rangle$ and $|XX\rangle$ (i.e. when the pulse area is increased) [21]. The two distinctive signatures of the two-photon excitation process are the slow increase of the intensity at low power and the change of the Rabi oscillations frequency with the pump power [15]. The slightly monotonous increase superimposed to the oscillations could be attributed to the slight chirp [22] of the excitation pulse (see supplementary) or by some incoherent contribution. Fed through the radiative cascade by the biexciton recombination, the exciton intensity is following a very similar dependence (Fig. 3b).

To evaluate the impact of this QD coherent control on the degree of entanglement of the emitted photon pairs, we performed quantum state tomography measurements under $\pi$-pulse excitation. By measuring the degrees of polarization correlation between the biexciton and the exciton in the linear, diagonal and circular basis, one can obtain a close approximation of the fidelity to the Bell state [23]. In Fig. 4, the six corresponding correlation histograms are displayed. In the linear and diagonal basis, they show a strong bunching when the polarizations of the exciton and the biexciton photons are parallel and an antibunching when they are orthogonal. This is the opposite in the circular basis where the antibunching occurs when the photons are prepared in parallel polarizations. A bunching can be observed when photons are projected in orthogonal circular polarizations. This set of correlation measurements shows a clear signature of entanglement. The polarization correlation contrast in a given basis $\mu$ is defined as [23]:

$$C_\mu = \frac{g^{(2)}_{xx,x}(0) - g^{(2)}_{xx,\bar{x}}(0)}{g^{(2)}_{xx,x}(0) + g^{(2)}_{xx,\bar{x}}(0)} \quad (1)$$

$g^{2}_{xx,x}(0)$ is the zero-delay second-order correlation between the first and second photons in the polarization basis and $g^{(2)}_{xx,\bar{x}}(0)$ is the cross-polarized second order correlation. Details about this measurements are provided in the supplementary. The respective contrasts extracted from them in the linear, diagonal and circular basis are:

$$\begin{aligned} C_{linear} &= 0.87 \pm 0.02 \\ C_{diagonal} &= 0.67 \pm 0.04 \\ C_{circular} &= -0.69 \pm 0.02 \end{aligned} \quad (2)$$

The fidelity of the state emitted by the quantum dot with respect to the expected state $|\psi^+\rangle$ is calculated from the three contrasts [19]:

$$f = \frac{1 + C_{linear} + C_{diagonal} - C_{circular}}{4} = 0.81 \pm 0.02 \quad (3)$$

This is a clear improvement compared to the fidelity of $0.72 \pm 0.01$ extracted from a previous quantum state tomography performed on the same quantum dot, using an above bandgap excitation [11]. This is related to the improvement of the coherence time consequent to the biexciton resonant excitation (see supplementary) [23].

**Figure 2** | Autocorrelation histograms, under resonant excitation, of **a** the excitonic photons, **b** the biexcitonic photons. Both data are acquired with the biexciton being resonantly excited with $\pi$-pulses. The data are presented without correction. The raw antibunching values were measured to $g^{(2)}_X(0) = 0.022 \pm 0.018$ and $g^{(2)}_{XX}(0) = 0.021 \pm 0.018$. Corrected from APDs dark count contribution, they reach: $g^{(2)}_X(0) = 0.004 \pm 0.018$ and $g^{(2)}_{XX}(0) = 0.003 \pm 0.018$.

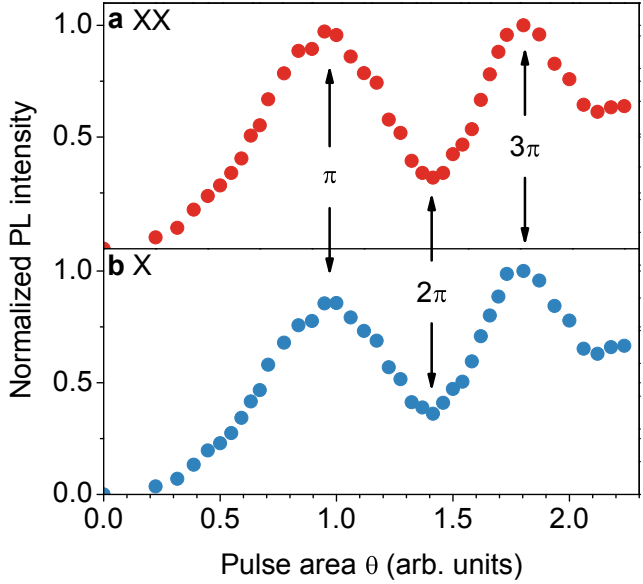

**Figure 3** | Integrated intensities, under resonant excitation of **a** the biexcitonic emission line, and of **b** the excitonic emission line versus the square root of the pump power (proportional to the excitation pulse area). The abscissa is renormalized in $\pi$-units such that the first biexciton intensity maximum, which corresponds to the first rotation from $|0\rangle$ to $|XX\rangle$ in the Bloch sphere, is reached for a pulse area equal to $\pi$. In the supplementary information, we evaluate that more than 75 percent of the excitation pulses generate a photon pair.

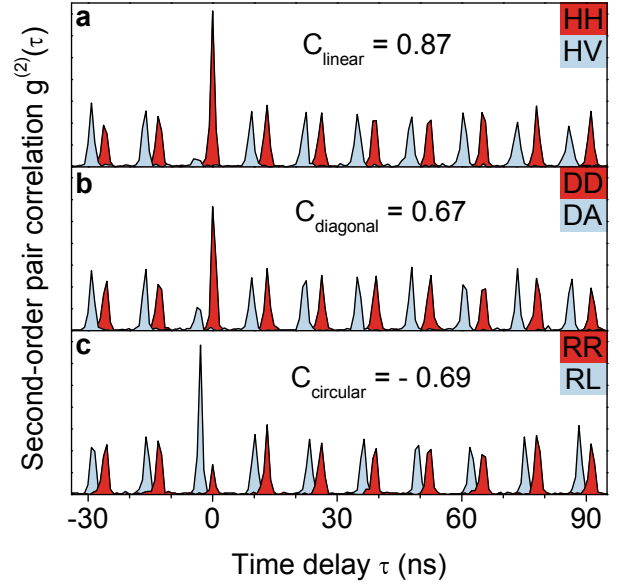

**Figure 4** | Biexciton-exciton polarization dependent cross-correlation histograms, under resonant excitation, in the **a** linear basis, **b** diagonal basis, and **c** circular basis. An antibunching is observed when the biexciton and exciton photons are projected in orthogonal polarizations in the linear and diagonal basis. A bunching is observed for parallel polarizations. In the circular basis, the opposite is observed. The relative areas of the zero delay peaks form these six histograms are used to derive the cross-polarization contrasts. For clarity, data corresponding to orthogonal polarization configurations are time-shifted.

In order to study the indistinguishability of the photons, two-photon interference measurements have been performed. The biexcitonic state of the quantum dot is excited twice every 13 ns by a pair of $\pi$-pulses separated by a 4 ns delay. Consequently, upon each excitation pulse the QD emits a pair of photons (X and XX). After spectral selection and projection of their polarization to the horizontal, the photons from the same transition (X and XX) were then fed into an unbalanced Mach-Zehnder interferometer (MZI) with a 4 ns path-length difference. Within the MZI the photons were prepared in orthogonal or parallel polarization making them distinguishable or indistinguishable with respect to their polarizations. The two outputs of this interferometer were detected by single-mode fiber-coupled single-photon counters and a histogram is built with the recorded coincidences (see Fig. 5). The two photons arrive at the beamsplitter with a delay of -8, -4, 0, 4, and 8 ns building a cluster of 5 peaks. Due to the 13 ns delay between the clusters, the two outer peaks of a cluster temporally overlap with corresponding peaks from the previous/successive cluster. However, the central peak, which reflects the two-photon interference probability, is well resolved and not affected by any overlap due to the 4 ns time separation between the pulses. This allows us a background free determination of the two-photon interference effect for both transitions (X and XX). Fig. 5a and 5b show a strong suppression of both coincidence peaks at zero delay when the two incoming photons are prepared in the same polarization (red and blue color) in contrast to the orthogonal polarization case (gray color). The quantitative evaluation of the raw data results in two-photon interference visibilities of $0.58 \pm 0.04$ for XX and $0.44 \pm 0.03$ for X. Taking into account the experimental imperfections such as avalanche-photodiodes (APDs) dark counts, the reduced mode overlap at the beamsplitter $(1-\epsilon = 0.95 \pm 0.01)$ and the residual two-photon emission probabilities, we obtain the corrected degrees of indistinguishabilities to be $0.84 \pm 0.05$ for XX and $0.69 \pm 0.04$ for X. The degree of indistinguishability can be independently checked with the relative intensities of the side peaks due to their known peak area ratios basically producing the same results: $0.86 \pm 0.03$ for XX and $0.71 \pm 0.04$ for X (for more details, see supplementary). The difference between the exciton and the biexciton visibilities arises from the fact that the biexciton state lifetime introduces a time jitter to the exciton population which results in a degradation of the excitonic photons indistinguishability [24–26]. We anticipate that this issue could be overcome by bringing a cavity mode in resonance with the biexciton transition. The enhancement of the biexciton radiative rate by a Purcell effect would efficiently reduce the undesired time jitter.

All the features presented here show that the resonant biexciton excitation is particularly suited for quantum computation or communication schemes [27] which require efficiently generated photons linked by a high degree of entanglement, and able to interfere correctly on a beam-splitter. Ultimately, more sophisticated coherent control of the quantum dot state could even further upgrade the performances of the source. The use of the pulse echo technique [28], suppressing the precession of the Bloch vector during the excitation, would be another step towards very pure states generation. Application of the TPE technique on symmetric quantum dots ($C_{3V}$) [29] with nearly zero fine structure splittings embedded in resonator [9] or nanowire structures [30] which allow ultra-high collection efficiencies would open the way to very competitive semiconductor entangled photon sources.

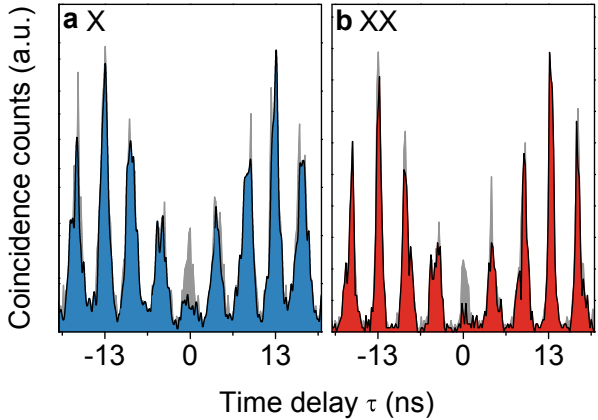

**Figure 5** | Two-photon interference histograms, obtained under resonant excitation for **a** the excitonic photons, **b** the biexcitonic photons. On both graphs, the colored plots correspond to the two-photon interference pattern of the parallel-polarized single photons whereas the gray graphs correspond to the two-photon interference of cross-polarized single photons. When photons arrive to the beam splitter in parallel polarization, i.e. they are indistinguishable, the central peak is strongly suppressed. The gray histograms are used to quantify the central peak suppression and to extract the visibilities of the two-photon interference. Side peaks were also used for cross-checking the extracted values. They were evaluated to $0.86 \pm 0.03$ for XX and $0.71 \pm 0.04$ for X after dark count and setup imperfection corrections.

**METHODS** The investigated sample consists of a single layer of self-assembled In(Ga)As quantum dots on a semi-insulating GaAs (001) substrate, and was grown by molecular beam epitaxy. The QDs are confined within a 260 nm thick GaAs layer, which acts as a lambda-cavity. A bottom distributed Bragg reflector (DBR) made up of 15 pairs of AlAs/GaAS (78.4 nm/65 nm) and one single pair of GaAs/AlAs on the top are grown to enhance the collection efficiency. To carry out $\mu$PL measurements the sample is mounted in a helium-flow cryostat at 4 K. An orthogonal geometry for the excitation and detection improves the rejection of scattered laser light. For the side-excitation with shaped pulses from a Ti:Sapphire laser system with a repetition rate of 76 MHz (see supplementary material for further information), the DBR structure acts as a waveguide. The QD emission is collected from the top and analyzed with a spectrometer/charge-coupled device combination. This can be either used to record the emission light spectrum with a resolution of 50 $\mu$eV or as a monochromator for spectral filtering. For the autocorrelation as well as for the indistinguishability measurements, fiber-coupled setups were used. To perform the quantum state tomography experiment, an assembly of waveplates ($\lambda/2$, $\lambda/4$) and polarizing beamsplitters project the quantum dot luminescence in the particular polarization basis before the spectral filtering of each individual emission line takes place. Three polarization independent interferential filters ($\Delta\lambda < 0.8\ meV$), adjusted on the laser wavelength, were mounted in the detection path for laser suppression. Furthermore, single-mode fibers, connected the APDs, provide a spatial filtering of the scattered laser light.

**Acknowledgements** The authors acknowledge L. Wang, A. Rastelli and O. G. Schmidt for providing the high-quality sample. The authors acknowledge financial support from the DFG via the project MI 500/23-1.

**Competing Interests** The authors declare no competing financial interests.

**Author contributions** M.M., S.B., K.D.J. and P.M. conceived and designed the experiments. M.M., S.B. and K.D.J. performed the experiments. M.M. and S.B. analyzed the data. S.B. and P.M. wrote the manuscript with input from the other authors.

**Correspondence** Correspondence and requests for materials should be addressed to S.B. and P.M.

# Supplementary information to "On-demand generation of indistinguishable polarization-entangled photons"

M. Müller†, S. Bounouar†⋆, K. D. Jöns and P. Michler⋆

**August 21, 2013**

*Institut für Halbleiteroptik und Funktionelle Grenzflächen, Universität Stuttgart, Allmandring 3, 70569 Stuttgart, Germany.*
†*These authors contributed equally to this work.* ⋆*e-mail: s.bounouar@ihfg.uni-stuttgart.de; p.michler@ihfg.uni-stuttgart.de*

## 1 Excitation-pulse characterization

In order to be strictly resonant with the virtual state and not with the individual exciton, and to avoid phonon-assisted processes, we reduced the spectral linewidth of the the excitation pulse from $500\,\mu eV$ to $95\,\mu eV$ by a simple pulse shaping technique. In a first step the laser pulse is diffracted by a $1200\,g/mm$ grating. It is then imaged on the $4\,\mu m$ core of a single-mode fiber which acts as a spatial spectral filter to reject undesired spectral components. Fig. 1a shows the resulting spectral shape obtained with this method. The central pulse-shaped laser energy can be tuned by a slight change of the horizontal position of the fiber coupling. Fig. 1b shows the temporal shape of the pulse. We measure a pulse length of $21.4\,ps$. Taking into account the spectral width of $FWHM_{pulse} = 95 \pm 1\,\mu eV$, we calculate a time-bandwidth product of $\tau_{pulse} \cdot FWHM_{pulse} = 0.492$. Therefore, the generated pulses are nearly transform-limited (Fourier-limit for Gaussian pulses: $\Delta t \cdot \Delta \nu = 0.441$) and contain just a slight frequency chirp, since the wavelength components are not re-arranged in time domain.

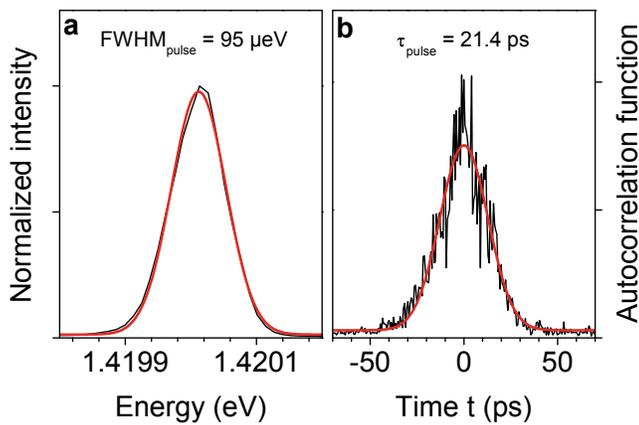

**Figure 1** | Excitation-pulse characterization. **a** spectral and **b** temporal shape of the excitation laser pulse (red lines: Gaussian fits).

## 2 Autocorrelation data analysis

The second-order autocorrelation measurements were carried out by using a fiber-coupled Hanbury Brown and Twiss setup. The correlation histograms are analyzed by integrating the peak areas, which are separated by the laser repetition pulse rate of 76 MHz. For $|\tau| > 0$ the peak areas are used to obtain the Poisson level for the normalization of $g^{(2)}(\tau = 0)$. Taking into account only the raw data without any background correction, we extract the following $g^{(2)}(0)$-values for the X and XX transitions, respectively:

$$g^{(2)}_{X,\mathrm{raw}}(0) = 0.0220 \pm 0.0177 \qquad (1)$$
$$g^{(2)}_{XX,\mathrm{raw}}(0) = 0.0206 \pm 0.0181$$

The data are then corrected from the background coincidences due to the APDs dark count rates. We evaluated this contribution by deriving the uncorrelated coincidences per channel:

$$N_c = (n_{s1} + n_{s2}) \cdot n_{dc} \cdot \tau_c \cdot T \qquad (2)$$

with $n_{s1}$ and $n_{s2}$ the count rates on the two APDs during the experiment ($n_{s1} = n_{s2} = 2500 - 3000\,\mathrm{cts./s}$ ($\sim 200$-$230\,\mathrm{kcts./s}$ at the first lense)), $n_{dc}$ the dark count rate ($n_{dc} = 250\,\mathrm{cts./s}$), $\tau_c$ the channel time width and $T$ the integration time. The subtraction of the detector noise reveals the following $g^{(2)}(0)$-values:

$$g^{(2)}_X(0) = 0.0037 \qquad (3)$$
$$g^{(2)}_{XX}(0) = 0.0031$$

The estimated statistical error exceeds the corrected $g^{(2)}(0)$-values. In order not to present unphysical error bars, we do not give them here.

## 3 Carrier lifetimes

Decay time experiments were performed on the exciton and biexciton transitions (Fig. 2). The measured curves are characteristic of a radiative cascade. The biexciton curve shows no rise time, except the one imposed by the 50 ps resolution of the APD, and is dominantly mono-exponential. The exciton shows a plateau at small delays, due to the biexciton recombination and decays with a biexponential dependence. A numerically resolved three-level model taking into account the ground state $|0\rangle$, the exciton state $|X\rangle$ and the biexciton state $|XX\rangle$ was used for the fit. In order to match the measured data, the numerical model was convoluted with the response function of the setup, given by the laser autocorrelation. Since the excitation process is a direct excitation of the biexciton, we used for initial conditions

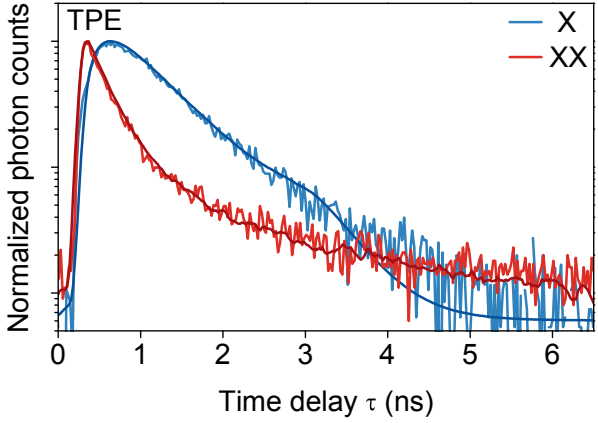
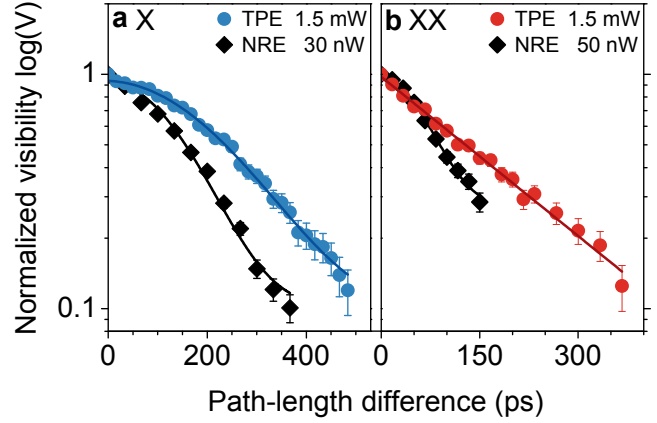

**Figure 2** | Time-resolved PL of the exciton (X) and biexciton (XX) after two-photon excitation including the numerical model described in the text.

**Figure 3** | Comparison of the first-order autocorrelation function for **a** the X and **b** the XX transitions in resonant two-photon excitation (TPE) and non-resonant excitation (NRE). The solid lines are the Gaussian and exponential fit-functions given in the text.

$n_{|0\rangle}(0) = 0$, $n_{|X\rangle}(0) = 0$, $n_{|XX\rangle}(0) = 1$. The fit parameters are the biexciton ($T_{1,XX}$) and the exciton ($T_{1,X}$) lifetimes. We extract $T_{1,XX} = 220\,\text{ps} \pm 20\,\text{ps}$ and $T_{1,X} = 400\,\text{ps} \pm 20\,\text{ps}$.

## 4 Estimation of biexciton preparation fidelity

Since the rotation angle in the Bloch sphere is proportional to the pulse area $\theta$, an increase of the pump power makes the quantum dot oscillate between the ground state and the biexciton state. Since the contrast of the first oscillation observed on Fig. 3 of the article is pronounced, the biexciton preparation fidelity is expected to be close to unity. Modeling the driving of a quantum dot with a chirped pulse requires a substantial theoretical study [1] and is out of the scope of this article. Thus, we provide a simple underestimation of the biexciton preparation fidelity. In the case of a power dependent exponential damping, with a transform-limited laser excitation, the amplitude of the oscillations in $\theta = 2\pi$ is smaller than in $\theta = \pi$. Thus by measuring the biexciton intensity at $\theta = 2\pi$ (which should be zero without damping), we can roughly estimate the loss of amplitude at $\theta = \pi$:

$$f_{XX}(\pi) > \frac{I_{XX}(\pi)}{I_{XX}(\pi) + I_{XX}(2\pi)} = 0.75 \quad (4)$$

This leads us to a biexciton preparation fidelity larger than 0.75.

## 5 Coherence time measurements

The first-order autocorrelation function is determined by self-interference measurements of the emitted photons on a beamsplitter of a Michelson interferometer. After a spectral filtering with a monochromator, the signal output is detected by a fiber-coupled APD. The interference fringe contrast $g^{(1)}(\tau)$ given by

$$g^{(1)}(\tau) = \frac{I_{max} - I_{min}}{I_{max} + I_{min}} \quad (5)$$

where $I_{max}$ and $I_{min}$ are the maximum and minimum intensities of the oscillating signal at the position $\tau$, is obtained by fitting a $sin^2$-function on the recorded data. Fig. 3 shows the comparison of the first-order autocorrelation function curves obtained under resonant excitation ($\pi$-pulse excitation) and under above bandgap excitation (excitation power close to saturation for both transitions, respectively). Gaussian as well as exponential shapes are observed for the individual $g^{(1)}(\tau)$-functions. To extract the coherence time $T_2$, the following fitting functions were used to analyze the data:

$$g^{(1)}_{\text{Gauss}}(\tau) \sim \frac{1}{\sqrt{2\pi}} \cdot \exp\left(-\frac{\pi}{2} \cdot \left(\frac{\tau}{T_2}\right)^2\right) \quad (6)$$

$$g^{(1)}_{\text{exp}}(\tau) \sim \exp\left(\frac{\tau}{T_2}\right) \quad (7)$$

Both graphs (Fig. 3a,b) show an improvement of the coherence time for the exciton (from 229 ps to 357 ps) and the biexciton (from 114 ps to 192 ps). This improvement is expected since resonant excitation introduces less decoherence for the emitter, and makes no use of phonons for carriers relaxation. For resonant and non-resonant excitation, the two curves corresponding to the exciton show a Gaussian shape (Fig. 3a). In both cases, the emitter is far from the linewidth imposed by the lifetime. Even for resonant excitation, some slow dephasing processes, such as spectral diffusion take place. The situation for the biexciton is more puzzling (Fig. 3b). For the non resonant case, the curve is again Gaussian, as it is the case for the exciton. However when the laser is set to resonance with the two-photon excitation, we observe for the biexciton a Lorentzian line shape. No clear evidence can be provided here to explain this change of line shape. Such a narrowing of the linewidth towards a Lorentzian spectrum with power or temperature has been observed in GaAs quantum dots. When the emitter energy fluctuates faster or with a smaller amplitude, this "motional narrowing" can occur [2]. It is unlikely that the fluctuations of the electronic environment surrounding the quantum dot get faster for the resonant excitation. This could be due to a reduction of their amplitude, since less traps are expected to be occupied with this pumping scheme.

## 6 Cross-polarization correlation measurements

By measuring the degrees of polarization correlation between the biexciton and the exciton in the linear, diagonal and circular basis, one can obtain a close approximation of the fidelity to the expected Bell state $|\psi^+\rangle$ [3]. The polarization correlation contrast in a given basis $\mu$ is defined as:

$$C_\mu = \frac{g^{(2)}_{xx,x}(0) - g^{(2)}_{xx,\bar{x}}(0)}{g^{(2)}_{xx,x}(0) + g^{(2)}_{xx,\bar{x}}(0)} \quad (8)$$

$g^2_{xx,x}(0)$ is the zero-delay second-order correlation between the first and second photons in the polarization basis and $g^2_{xx,\bar{x}}(0)$ is the cross-polarized second order correlation. The corresponding, detector noise subtracted (see Eq.2 with 3000 cts./s on one APD), $g^2_{xx,x}(0)$ values measured in the three polarization basis are:

$$g^{(2)}_{H,H}(0) = 2.40 \pm 0.02 \quad g^{(2)}_{H,V}(0) = 0.17 \pm 0.01$$
$$g^{(2)}_{D,D}(0) = 2.18 \pm 0.04 \quad g^{(2)}_{D,A}(0) = 0.43 \pm 0.03 \qquad (9)$$
$$g^{(2)}_{R,R}(0) = 0.31 \pm 0.01 \quad g^{(2)}_{R,L}(0) = 1.70 \pm 0.02$$

The respective contrasts extracted from the linear, diagonal and circular basis are:

$$C_{linear} = 0.87 \pm 0.02$$
$$C_{diagonal} = 0.67 \pm 0.04 \qquad (10)$$
$$C_{circular} = -0.69 \pm 0.02$$

The fidelity of the state emitted by the quantum dot with respect to the expected state $|\psi^+\rangle$ is calculated from the three contrasts [3]:

$$f = \frac{1 + C_{linear} + C_{diagonal} - C_{circular}}{4} = 0.81 \pm 0.02 \qquad (11)$$

## 7 Two-photon interference experiment

The two-photon interference pattern is composed of five peaks clusters separated by the repetition time of the laser source. The five peaks, from left to right, correspond to the cases where the two photons arrive at the second beamsplitter with a time delay of $-2\,\delta t, -\delta t, 0, \delta t$ and $2\,\delta t$, respectively, where $\delta t$ is the fixed delay (4 ns) of the Mach-Zehnder interferometer (see Fig. 4). For distin-guishable photons with different polarizations, the expected peak area ratio in the zero delay cluster is $1:2:2:2:1$ [4]. For the delayed clusters, the peak ratio is $1:4:6:4:1$. The 4 ns delay is introduced for coincidence peaks separation between the two arms of the unbalanced Mach-Zehnder interferometer. Due to the 13 ns repetition rate of the laser we observe an overlap of the fourth and fifth ($B_L$) peak of the delayed pattern with the first ($A_1$) and second ($A_2$) peak of the zero delay pattern (Fig. 5a). Fig. 5b shows the two-photon interference pattern acquired with the biexciton photons in orthogonal polarization. The five peaks correspond to the envelope of the laser histogram in the graph above. In particular the mentioned overlap of the zero delay pattern with the delayed clusters is obvious. Since the degree of indistinguishability is given by the ratio between the middle peak $A_3$ and the half of the sum of the side-peaks $A_2 + A_4$ in Fig. 5a, we have to take into account the overlap of the side peaks $A_2$ and $A_4$ with the peaks $B_L$ and $B_R$. That is, the area of peak $A_2^*$ is the sum of peak $A_2$ and the fifth peak $B_L$ of the next cluster. To recalculate the exact integral of the peaks $A_2$ and $A_4$, we take advantage of the well known peak area ratio for each individual cluster. Taking into account the ratio $A_2/A_1 = A_2/B_L = 2$ we can write the peak areas $A_2^*$ and $A_4^*$ as a function of $A_2$ and $A_4$:

$$A_2^* = A_2 + B_L = A_2 + A_2/2$$
$$= 3/2 \cdot A_2 \qquad (12)$$

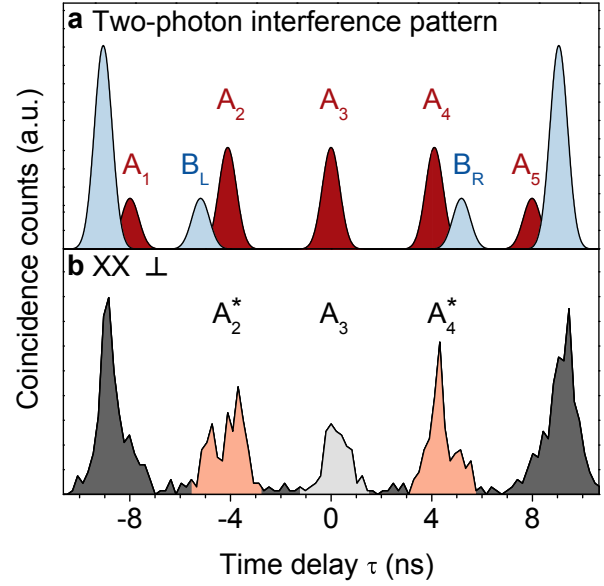

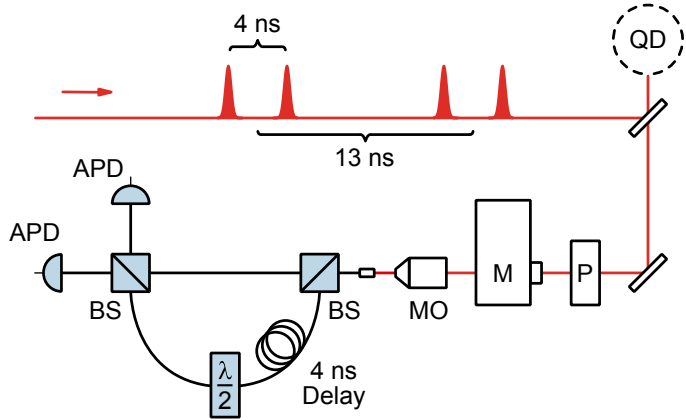

**Figure 4** | Scheme of the two-photon interference experiment. A delay stage is used to generate a pair of excitation laser pulses with a time distance of 4 ns and a repetition rate of 13 ns. Therefore, the QD is excited twice within the time delay of the laser pulses. The emitted photons are first projected in a defined linear polarization (polarizer P) and afterwards spectrally filtered by a monochromator M. A microscope objective MO (NA = 0.4) couples the light into a fiber-based Mach-Zehnder interferometer with a fixed fiber delay of $\delta t = 4$ ns in one arm. Additionally, a $\lambda/2$-plate is used in this path for cross-polarization measurements. Fiber-coupled APDs detect the two outputs of the second beamsplitter and a histogram is built with the recorded coincidences.

**Figure 5** | **a,** Scheme of the two-photon interference coincidence pattern. Because of the excitation pulse delay of 4 ns and the repetition rate of 13 ns, peaks from different excitation clusters are overlapping. **b,** Two-photon interference histogram for distinguishable biexciton photons (cross-polarized $\lambda/2$ position). The peaks $A_2^*$ and $A_4^*$ are compositions of the peaks $A_2$, $B_L$ and $A_4$, $B_R$, respectively. Since the ratio of the peak areas is known for each individual cluster, we can recalculate the visibility from the histogram (see explanation in the text).

The same way,

$$A_4^* = A_4 + B_R = A_4 + A_4/2$$
$$= 3/2 \cdot A_4 \qquad (13)$$

The visibility V can be defined as a function of the peak areas $A_2$, $A_3$ and $A_4$ [4]:

$$V = 1 - \frac{A_3}{(A_2 + A_4)/2} \qquad (14)$$

We introduce Eq. 12 and Eq. 13 in Eq. 14 to give an expression for the visibility as a function of the peak areas $A_2^*$ and $A_4^*$:

$$V = 1 - \frac{A_3}{(A_2^* + A_4^*)/3} \qquad (15)$$

| Visibilities | X | XX |
|---|---|---|
| Raw data | $0.46 \pm 0.02$ | $0.59 \pm 0.02$ |
| APD corrected | $0.63 \pm 0.02$ | $0.77 \pm 0.03$ |
| APD and BS corrected | $0.71 \pm 0.04$ | $0.86 \pm 0.03$ |

**Table 1** | Indistinguishability visibilities for a raw data analysis and for a correction of the detector noise (APD corrected) as well as for the correction of the non-perfect beamsplitter overlap and the residual two-photon emission of the photons (BS corrected).

| Visibilities | X | XX |
|---|---|---|
| Raw data | $0.44 \pm 0.03$ | $0.58 \pm 0.04$ |
| APD corrected | $0.61 \pm 0.03$ | $0.75 \pm 0.04$ |
| APD and BS corrected | $0.69 \pm 0.04$ | $0.84 \pm 0.05$ |

**Table 2** | Indistinguishability visibilities obtained by using Eq. 16 for data analysis with the different corrections (see text and Tab. 1).

The results for different corrections applied on the raw data is listed in Tab. 1. For the first correction, we subtracted the detector noise evaluated with Eq. 2 (APD correction). Additionally, the indistinguishability of the photons is degraded by the beamsplitter mode overlap imperfection $(1 - \epsilon = 0.95)$ on the fiber-coupled beamsplitter and the residual two-photon emission (0.004 for the exciton and 0.003 for the biexciton, respectively). This factors are also considered to correct the visibility (BS correction) [4]. The characterization of the two-photon interference setup reveals a perfect 50/50 reflection/transmission ratio of the second beamsplitter. Therefore the beamsplitter ratio does not influence the data correction analysis.

We also derived the visibility with the help of the cross polarized histogram Fig. 5b and checked that the results were consistent with the previous analysis. After renormalization, on can deduce it from the zero delay peaks areas respectively in cross- ($g^{(2)}_\perp(0)$) and in parallel polarization ($g^{(2)}_\parallel(0)$):

$$V = 1 - \frac{g^{(2)}_\parallel(0)}{g^{(2)}_\perp(0)} \qquad (16)$$

The results obtained after the different corrections described above are listed in Tab. 2.